\documentclass[a4paper,conference]{IEEEtran}

\usepackage{times}
\usepackage{url}
\usepackage{latexsym}
\usepackage{mathtools}
\usepackage{eqnarray,amsmath} 
\usepackage{amsfonts}
\usepackage{float}
\usepackage{xcolor}
\usepackage{subfig}
\usepackage{float}
\usepackage{enumerate}
\usepackage{etoolbox}
\usepackage{booktabs}
\usepackage{graphicx}

\begin{document} 
	\title{End-to-end Triplet Loss based Emotion Embedding System for Speech Emotion Recognition}

	\author
	{
		\IEEEauthorblockN
		{	Puneet Kumar\IEEEauthorrefmark{2},			 	
			Sidharth Jain\IEEEauthorrefmark{3}, 			
			Balasubramanian Raman\IEEEauthorrefmark{2},		
			Partha Pratim Roy\IEEEauthorrefmark{2} and 		
			Masakazu Iwamura\IEEEauthorrefmark{4} 			
		}
		\IEEEauthorblockA
		{\IEEEauthorrefmark{2} Department of Computer Science \& Engineering
		}
		\IEEEauthorblockA
		{\IEEEauthorrefmark{3}Department of Civil Engineering
		}
		\IEEEauthorblockA
		{Indian Institute of Technology Roorkee, India
		}
		\IEEEauthorblockA
		{\IEEEauthorrefmark{4}Dept. of Computer Science and Intelligent Systems, Osaka Prefecture University, Japan	
		}
		\IEEEauthorblockA
		{ {\tt \{pkumar99@cs, sjain@ce, bala@cs, partha@cs\}.iitr.ac.in, masa@cs.osakafu-u.ac.jp}
		} 
	}
	
	\maketitle
	
	\begin{abstract}
		In this paper, an end-to-end neural embedding system based on triplet loss and residual learning has been proposed for speech emotion recognition. The proposed system learns the embeddings from the emotional information of the speech utterances. The learned embeddings are used to recognize the emotions portrayed by given speech samples of various lengths. The proposed system implements Residual Neural Network architecture. It is trained using softmax pre-training and triplet loss function. The weights between the fully connected and embedding layers of the trained network are used to calculate the embedding values. The embedding representations of various emotions are mapped onto a hyperplane, and the angles among them are computed using the cosine similarity. These angles are utilized to classify a new speech sample into its appropriate emotion class. The proposed system has demonstrated 91.67\% and 64.44\% accuracy while recognizing emotions for RAVDESS and IEMOCAP dataset, respectively.
	\end{abstract}


	\IEEEpeerreviewmaketitle
	
	\section{Introduction}\label{sec:introduction}
	The need to develop efficient speech processing systems that are capable of recognizing various emotions from the speech is increasing at a fast rate~\cite{zeng2009survey}. Such systems are useful for a wide range of applications such as robotics, security, service delivery, language translation, automated identification, intelligent toys, and lie detection~\cite{el2011survey}. Speech is one of the important ways for a human to portray complex emotions. It can also be used as an efficient method of human-machine interaction~\cite{vogt2008automatic}. However, a major challenge in human-machine interaction is the correct detection of emotion from speech. It is natural for a human to recognize underlying emotions during their spoken interactions. However, it is difficult for machines to recognize complex emotions in natural speech.\vspace{.02in}
	
	Emotional information included in a speech signal depends on several factors such as speaker, style, language, gender, accent, and sample-duration~\cite{vogt2008automatic}. The notions of various emotions are highly subjective. People interpret them differently depending upon their culture and environment. Likewise, labeling the speech data with suitable emotion during its preparation is also subjected to human variability. A potential approach to reduce human fluctuations in Speech Emotion Recognition (SER) is to develop an SER system that can recognize speech corresponding to various emotions without human intervention. Such systems are called end-to-end systems, and as opposed to the conventional methods of emotional speech recognition, they do not require manual crafting of acoustic features. There is a need to develop an end-to-end SER system that can learn the emotional patterns in input speech data despite the aforementioned variations and bypassing the intermediate steps of speech processing~\cite{chiriacescu2009automatic}.\vspace{.02in} 
	
	The proposed model is based on Residual Neural Network (ResNet) architecture. The model is trained using softmax pre-training and triplet loss function. Then the embedding values are calculated from the weights between the fully connected and embedding layers. The embeddings are mapped onto a hyperplane, and cosine similarity values are calculated by measuring the cosine of the angles among the embedding representations of various emotions. Smaller the angle, higher will be the cosine similarity. The model is trained to minimize the angles among the representations of the speech of similar emotions and maximize the angles among the representations of the speech of different emotions. The computed angles are utilized to classify a new speech sample into its appropriate emotion class. The proposed approach has been validated for two emotional speech datasets - The Ryerson Audio-Visual Database of Emotional Speech and Song (RAVDESS) and The Interactive Emotional Dyadic Motion Capture (IEMOCAP). Recognition accuracies of 91.67\% and 64.44\% have been observed for RAVDESS and IEMOCAP dataset, respectively.\vspace{.02in}
	
	The major contributions of the paper are as follows. Firstly, a deep neural end-to-end SER system based on triplet loss and residual learning has been proposed. The proposed system is capable of learning emotion-related information from a labeled emotional speech dataset in the form of embeddings. Secondly, the embeddings learned by the proposed system are used to classify the speech samples of various lengths into appropriate emotion classes. Using the embeddings, the proposed system can estimate the emotions in unseen speech utterances. \vspace{.02in}
	
	The rest of the paper is organized as follows. Existing work in the context of speech emotion recognition has been surveyed in Section~\ref{sec:lr}. Section~\ref{sec:proposed} elaborates on the proposed methodology. In Section~\ref{sec:implementation}, implementation details and experimental results have been discussed. Finally, Section~\ref{sec:conclusion} concludes the paper and highlights the scope for future research.	
	
	\section{Related Work}\label{sec:lr}
	In recent years, several SER approaches	have been developed. Feature-based speech recognition systems attempt to extract characteristics from acoustic features such as fundamental frequencies, pitch, prosody, voice quality, Mel frequency cepstrum coefficient (MFCC), and linear prediction cepstrum coefficient (LPCC). They use these features to analyze the emotion patterns of speech samples~\cite{el2011survey}. In this context, C. Lee et al.~\cite{lee2005toward} used pitch, formants, and speech rate to differentiate positive emotions from negative emotions in speech signals. In another work, J. Rong et al.~\cite{rong2009acoustic} developed a data pre-processing technique to extract the most relevant acoustic features for emotion recognition. It has been observed that the features derived from high-key emotions such as happiness, anger, and interest show similar properties among themselves which are very different from low-key emotions such as sadness and despair. Hence, there is a need to derive an SER method that is independent of the polarity of the emotional features.\vspace{.02in}
	
	Hidden Markov model (HMM) based statistical methods and support vector machine (SVM) based classifiers have also been used for SER. For example, J. Lorenzo et al.~\cite{lorenzo2015emotion} proposed an HMM-based method to detect and alter the emotional context of a speech sample while preserving the identity of the speaker. In another work, P. Shen et al.~\cite{jain2018cubic} trained an SVM based classifier to differentiate the emotions present in speech signals based on acoustic features. They used SVM to detect speech emotions accounting for gender-based variations. One of the major problems with using HMM-based SER models is that they are not always able to reliably estimate the parameters of global speech features~\cite{el2011survey}.\vspace{.02in}
	
	The above-mentioned approaches to emotional speech recognition require manual crafting of acoustic features. Hence, it is challenging to come up with an end-to-end SER system using them~\cite{lecun2015deep}. Neural network based models are capable of automatically extracting the features from the training data. In context of using them for SER, Stuhlsatz et al.~\cite{stuhlsatz2011deep} compared the performance of a neural network based Discriminant Analysis with SVM for the classification of emotional speech utterances. The neural network based classifier was observed to outperform the SVM based classifier for speech emotion detection. In another work, Mao et al.~\cite{mao2014learning} used CNN to extract the features from speech spectrograms. Then they classified the features using a binary classifier. Their model outperformed classic machine learning models.\vspace{.02in}
	
	Deep learning based systems have been used for other speech processing tasks as well. For instance, speaker recognition has been implemented by training speaker embeddings and differentiating the speakers based on them~\cite{li2017deep}. Similarly, A. Jain et al.~\cite{jain2018improved} implemented speaker independent accent embedding to differentiate multi accent speech. In the context of RNN based SER, N. Majumder et al.~\cite{majumder2019dialoguernn} implemented attention-based RNN to keep track of the identities of the speakers portraying specific emotions through conversations. In another work, S. Sahoo et al.~\cite{sahoo2019segment} utilized a pre-trained deep convolutional neural network to predict the emotion classes of the audio segments. Deep neural networks along with residual learning~\cite{he2016deep} and triplet loss~\cite{schroff2015facenet} are commonly used for facial expression recognition. They have been used in the area of speech processing also. For example, J. Kim et al.~\cite{kim2017deep} implemented a deep residual network for speech emotion recognition. In another work, H. Bredin~\cite{bredin2017tristounet} used triplet loss along with LSTM to learn the embeddings for speech sequences. The embeddings were later used for speaker identification. As speech emotion recognition is a counterpart of facial emotion recognition, triplet loss and residual learning based techniques may prove to be useful for SER also.\vspace{.02in} 
	 
	The success of deep neural networks for various speech processing tasks advocates their suitability for SER. However, end-to-end SER using triplet loss and residual learning along with deep neural networks has not been explored to its full potential. With that as an inspiration, various state-of-the-art deep neural architectures have been implemented and the best performing one is implemented in the proposed work. The proposed system also overcomes the issues with existing SER approaches, i.e., need for manual crafting of acoustic features, bias towards the polarity of the emotional features in feature-based SER, and unreliability of statistical SER systems in estimating the parameters of the global speech features.
	
	\section{Proposed system}\label{sec:proposed}
	This section describes an end-to-end approach to learn the embedding representations from emotional speech and use them for speech emotion recognition.	
	\subsection{Problem Formulation}
	Consider $d$-dimensional space $\mathbb{R}^d$ where elements in $\mathbb{R}^d$ are represented as $\{x_1, x_2, x_3,..,x_d\}$ where $x_i$ is a $d$-tuple that denotes an embedding vector $f(x) \in \mathbb{R} ^d$ mapped from a set of speech utterances $\sum_{j}^{} y_j$. The projections of such embedding vectors are represented in a hyperplane where emotion similarity is measured using cosine similarity. The objectives of the proposed technique are to:
	\begin{enumerate}[i]
		\item 
		Learn the embeddings from input speech utterances,  
		\item 
		Visualize the embeddings projected in a hyperplane to analyze the learned emotion patterns, 
		\item
		Use the learned embeddings to classify an unseen speech utterance into an appropriate emotion class.
	\end{enumerate} 
	
	\begin{figure}[]
		\begin{center}
			\captionsetup{justification=centering}
			\includegraphics[width=0.45\textwidth]{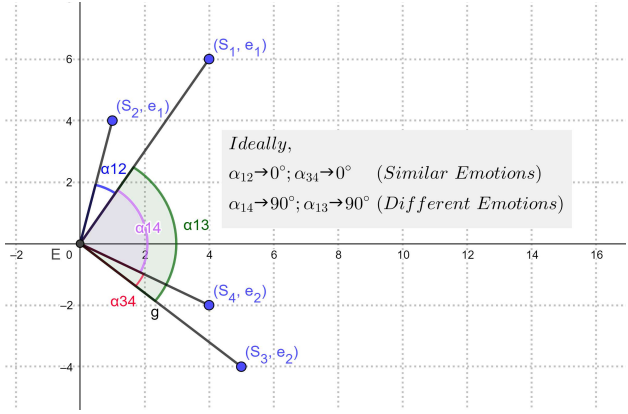}	
			\caption{Hyperplane with projections of speech utterances.\vspace{-.15in}} 
			\label{fig:prob}
		\end{center}
	\end{figure} 
	
	Fig.~\ref{fig:prob} explains the core hypothesis of the proposed work. Ideally, the emotion recognition should be independent of the length of the speech utterance. For example, the speech utterance $S1$ is longer than $S2$, but both incorporate the same emotion $e1$. Hence, the angle $\alpha_{12}$ between them is expected to be smaller than the angles between the speech emotions of dissimilar emotions such as  $\alpha_{13}$ and  $\alpha_{14}$.

	\subsection{Methodology}
\begin{figure}[]
		\begin{center}
			\captionsetup{justification=centering}
			\includegraphics[width=0.45\textwidth]{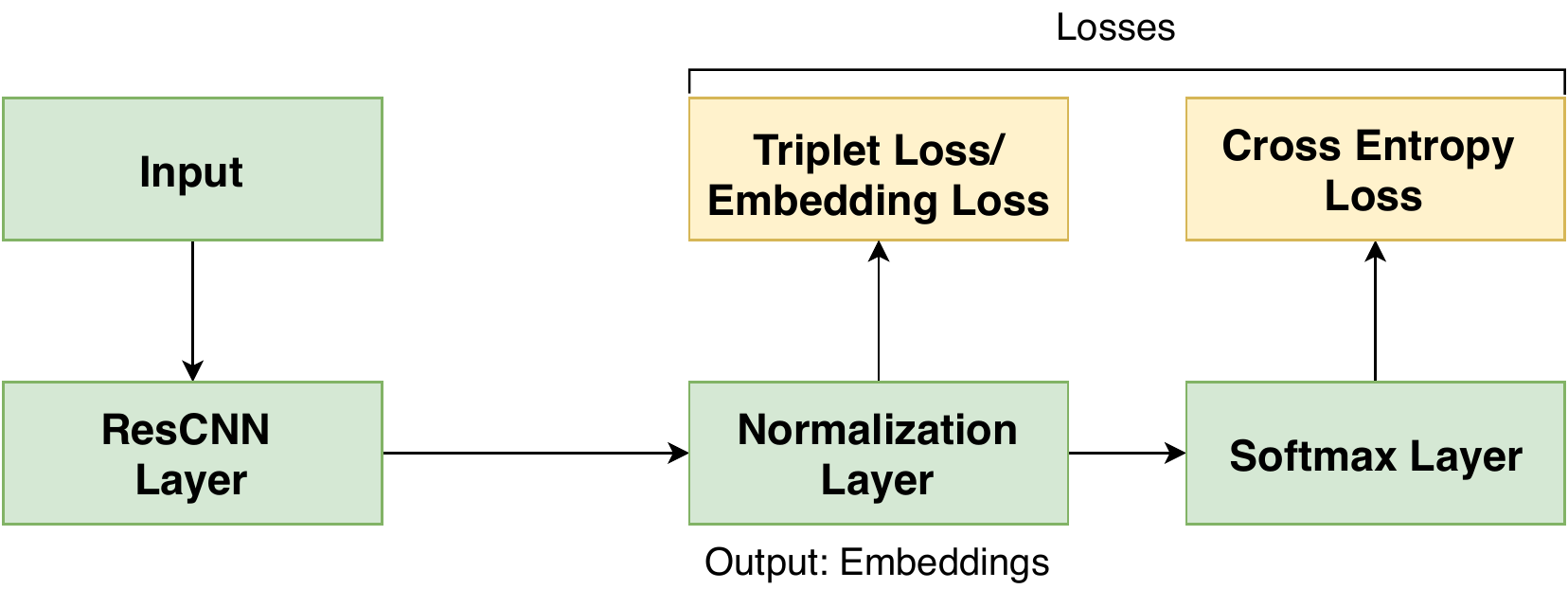}	
			\caption{Representation of the proposed methodology.\vspace{-.2	in}} 
			\label{fig:proposed}
		\end{center}
	\end{figure}

	Various phases of the proposed methodology are represented in Fig.~\ref{fig:proposed} and discussed in the following sections.
	
	\subsubsection{Phase I: Initialization and Pre-processing}
	The embeddings are initialized and data is pre-processed.
	
	\subsubsection*{a. Emotion Embedding Initialization} 
	Emotion embedding is a technique to represent the emotional information of the speech in the form of vectors. It learns the emotional content of the speech and constructs the vector representations for it. The network weights are defined by a temporal average layer which takes frame-level activations and computes the layer activation $h$ as follows:
	{\fontsize{9.5}{12}\selectfont 
	\begin{eqnarray}
		\label{eq:eq1}
		\begin{split}	
		&\ \ \ h= \frac{1}{T} \sum_{t=0}^{T-1} x(t)\\ 		
		&\begin{cases}
		where:\\
		T:\ number\ of\ frames\ in\ a\ speech\ utterance.\\
		t:\ time\ instance.\\
		x:\ embedding\ at\ given\ time\ instance\ t.\\
		h:\ activation\ of\ the\ layer.\\
		\end{cases}\\
		\end{split}
	\end{eqnarray}
	} \vspace{-.1in}		
	
	Emotion embeddings are learned using the weights between the fully connected layer and the embedding layer after the training completes. Here, initial model weights are initialized using softmax pre-training, and the final weights are assigned to these embeddings in later phases.  
	
	\subsubsection*{b. Audio Cache Generation \& Pre-processing}	
	The emotional speech dataset contains audio-clips with emotion labels. For each emotion label, the data is divided into a training set and a testing set. The cache is generated for both, which involves sampling the audio files and trimming the silence. After generating the audio cache, MFCC windows are randomly sampled from it. Since the model has been implemented for constant input shape, an appropriate value for the number of windows for the input signal had to be tuned. We took 10 windows from each input sample using random sampling of 10 continuous MFCC frames. For each window, 39 input values are present which correspond to 13 MFCC values, 13 first-derivative values, and 13 second-derivative values. If the number of windows for an input signal was less than 10, then zero padding was used to keep the input size fixed.\vspace{.1in}
	
	\subsubsection{Phase II: Embedding Training}\label{sec:II}
	A fully connected layer projects the utterance-level representations as embeddings. Emotion characteristics of the speech are learned by training embedding vectors for each emotion. The cross-entropy loss function is used to train the network along with triplet loss.
	
	\begin{itemize}
		\item \textbf{Cross-entropy Loss (Softmax Loss)}:
		When the output probability of a classification model is between 0 and 1, the cross-entropy loss function can be used to measure its performance. It produces stabler convergence for the data with noisy labels than other methods like mean absolute error loss, and categorical cross-entropy loss~\cite{zhang2018generalized}. \vspace{.05in}
		
		\item \textbf{Triplet Loss}:
		It is an optimization approach that compares a baseline input to a positive input and a negative input. It takes three speech samples and compares them in pairs. The distance between baseline input and positive input is minimized, and the distance between the baseline input and negative input is maximized~\cite{schroff2015facenet}. 
				
		\end{itemize}	
		{\fontsize{9.5}{12}\selectfont 
		\begin{eqnarray}
			\label{eq:eq2}
			\begin{split}	
				\ \ \ \ \ \ &J= \sum_{i=0}^{N} F(x_i^{e_1}, x_i^{e_2}, x_i^{e_3}) \\
				&F(e_1, e_2, e_3) = \\&max\{d(f(e_1),f(e_2)) - d(f(e_3),f(e_2)) + \alpha , 0\} \\		
			&\begin{cases}
				where:\\
				J: triplet\ loss\ cost\ function.\\
				F: intermediate\ function.\\
				x_i^{e_1}:\ embeddings\ for\ emotion\ 1.\\
				x_i^{e_2}:\ embeddings\ for\ emotion\ 2.\\
				x_i^{e_3}:\ embeddings\ for\ emotion\ 3.\\
				d(a,b): distance \ between\ point\ a\ and\ point\ b.\\
				f:\ mapping\ function\ for\ the\ embeddings.\\
				\alpha:\ margin\ between\ emotion\ embedding\ pairs.\\
			\end{cases}
			\end{split}
			\end{eqnarray}
		} \vspace{-.1in}
	
 	Triplet loss function, $J$ is defined for three emotion embeddings: emotion 1, emotion 2, and emotion 3. Assuming that emotion 1 is similar to emotion 2, but it is dis-similar to emotion 3. Then the distance between $e_1$ and $e_2$ will be minimized, and that between $e_1$ and $e_3$ will be maximized. This is achieved by function $F$. Triplets of various emotion pairs are chosen at random, and all the speech utterances are covered. The cost function $J$ computes the overall triplet loss.\vspace{.1in}
 	
	\textbf{Cosine Similarity}: Triplet loss is internally optimized using cosine similarity. The cosine similarity has successfully been used to check the similarity among the texts of unequal lengths. With that inspiration, the proposed method has implemented it to check the similarity among speech samples of varying lengths. It calculates the cosine of the angle between the vectors projected in a multi-dimensional space. Smaller the angle becomes, higher will be the cosine similarity. Triplet loss checks the embeddings in pairs. It aims to maximize the cosine similarities for the pairs with the same emotions and minimize those with different emotions	
	{\fontsize{9.5}{12}\selectfont 
	\begin{eqnarray}
	\label{eq:eq3}
	\begin{split}	
	&\ \ \cos(x_i, x_j) = x_i^Tx_j \\
	&\begin{cases}
	where:\\
	x_i\ and\ x_j:\ two\ emotion\ embeddings.
	\end{cases}\\
	\end{split}
	\end{eqnarray}
	} 	 
		
	The training process utilizes the above-mentioned concepts, i.e., softmax loss, triplet loss, and cosine similarity. It is carried out in the following two phases.\vspace{.02in}

	\subsubsection*{a. Softmax Pre-training}
	Softmax pre-training computes both softmax loss and triplet loss, and it trains the model for softmax loss. During pre-training, softmax pre-training is used to initialize the weights of the network. It maps the non-normalized output of the network to a probability distribution over predicted output classes. Softmax pre-training has been found to help avoid getting stuck in a local minimum and producing stabler convergence along with triplet loss~\cite{li2017deep}.\vspace{.02in}
	
	\subsubsection*{b. Embedding Training with Triplet Loss}
	This is the full training phase. It computes and trains on both loss values, i.e., softmax loss and triplet loss. Before generating the embeddings, the model weights are $L_2$ normalized. Then the embedding vector is generated, and triplet loss is calculated and minimized. Although the triplet loss function is well-used for face emotion recognition, it has also found its applications in the area of speech recognition where it has been used to learn and represent the speaker embeddings from speech utterances~\cite{li2017deep}. Softmax cross-entropy loss works well for a fixed number of classes. However, when there are a variable number of output classes then triplet loss can be used to learn good embeddings for each variation of each class~\cite{schroff2015facenet}.\vspace{.02in}
	
	It is observed that the softmax pre-training along with cross-entropy loss resulted in better performance than triplet loss implementation along with cross-entropy loss. The combination of triplet loss implementation along with cross-entropy for the input pre-processed with softmax pre-training performed even better. \vspace{.05in}

	\subsubsection{Phase III: Emotion Inference}		
	The steps to infer the emotion category for the unseen samples are as follows. 
	\begin{enumerate}[a.]
		\item
		Compile the test speech utterances in a new folder. 
		\item 
		Update the cache for the new folder. 	
		\item 
		Generate new embeddings using the trained network. 
		\item 
		Check cosine similarity for the generated embeddings in comparision with already generated embeddings for the emotional classes in the training phase.  
		\item 
		Project various embeddings onto the $\mathbb{R}^d$ hyperplane and measure the angles among them to determine their emotion category.
	\end{enumerate}

	The model training focuses on learning the emotion embeddings using triplet loss and cosine similarity. The cosine similarity is used for the inference as well. It helped to learn a good representation of the emotional information in speech utterances and gave an idea about the quality of the embeddings produced in the testing phase.
	
	\section{Experiments and results}\label{sec:implementation}	
	This section discusses the experimental implementation and analyses the results.
	\subsection{Implementation}	
	\subsubsection{Experimental Setup}
	Model training has been performed on Nvidia RTX 2070 GPU with 2304 CUDA cores, 288 Tensor cores, and 8 GB Virtual RAM. Model testing has been carried out on Intel(R) Core(TM) i7-7700, 3.70 GHz, 16GB RAM CPU system with Ubuntu 18.04.\vspace{.02in}
	
	\subsubsection{Dataset Selection and Training Strategy}\label{subsec:dataset} 
	Speech emotion recognition experiments have been conducted on the following datasets.	
	
	\begin{itemize}
		\item
		\textbf{RAVDESS}~\cite{livingstone2012ravdess} - Ryerson Audio Visual Data of Emotional Speech and Song (RAVDESS) is a standard scripted speech dataset containing eight emotion labels: happy, calm, sad, angry, surprise, fear, neutral, and disgust. Two sentences have been spoken and sung by 24 different speakers. It contains total of 7356 samples.\vspace{.02in}
		
		\item	
		\textbf{IEMOCAP}~\cite{busso2008iemocap} - Interactive Emotional Dyadic Motion Capture (IEMOCAP) is a relatively complex dataset containing impromptu and acted speech samples recorded using emotion sensitive human computer interfaces. It covers 10039 samples labeled with nine different emotion labels. Ten speakers have annotated a total of 12 sentences.	
	\end{itemize}   

	Final implementation has been carried out for complete RAVDESS and IEMOCAP datasets using 70\%-30\% training-testing split and 10-fold cross-validation. The model training contains two phases: softmax pre-training and triplet loss training. Softmax pre-training has been carried out for 1000 epoch, followed by 1000 epoch of triplet loss fine-tuning. Early stopping has been used to avoid overfitting. The model training is stopped when the accuracy does not improve by at least 0.0005 within 15 epochs. Linear decreasing learning rate ranged from 0.05 to 0.005 has been used. The results have been discussed in Section~\ref{sec:results}.\vspace{.05in}
	
	\subsubsection{Network Architecture}\label{subsec:archi}
	The ablation study to choose the appropriate network architecture has been performed with a sliced RAVDESS dataset containing 1600 samples, with 200 samples for each emotion class. The dataset was divided into training and testing sets in 70\% and 30\% ratio. Fully Connected (FC) network, CNN, Residual Neural Network (ResNet), RNN, Long Short Term Memory (LSTM) based RNN, and Gated Recurrent Unit (GRU) based RNN have been evaluated. They have been analyzed to extract emotional features embedded in acoustic input, train utterance level emotion embeddings, and train the network using cosine similarity based triplet loss function. Their details have been presented in Table~\ref{tab:ablation}. Here `x' represents the total number of layers. The analysis is performed for x = 6 to 15.\vspace{.02in} 

	\begin{table*}[]
		\centering
		{\fontsize{8}{10}\selectfont	
		\caption{Summary of the ablation study}\vspace{-.15cm}
		\label{tab:ablation}
		\resizebox{.6\textwidth}{!}
			{
			\begin{tabular}{@{}llcc@{}} \toprule
				\textbf{Architecture} & \textbf{Details}                           & \textbf{x} & \textbf{Accuracy} \\ \midrule
				FC-x                  & Fully Connected network with x layers      & 7          & 47.22\%           \\
				CNN-x                 & Convolutional Neural Network with x layers & 8          & 58.33\%           \\
				RNN-x                 & Simple x-layered Recurrent Neural Network  & 8          & 55.56\%           \\
				LSTM-x                & x-layered RNN with LSTM units              & 7          & 56.94\%           \\
				GRU-x                 & x-layered RNN with GRU units               & 8          & 54.16\%           \\
				\textbf{ResNet-x}     & Residual Neural Network with x layers      & 11         & \textbf{61.11\%}  \\ \bottomrule
			\end{tabular}%
			}
		} 
	\end{table*}
	
	\begin{figure}[]
		\begin{center}
			\captionsetup{justification=centering}
			\includegraphics[width=0.35\textwidth]{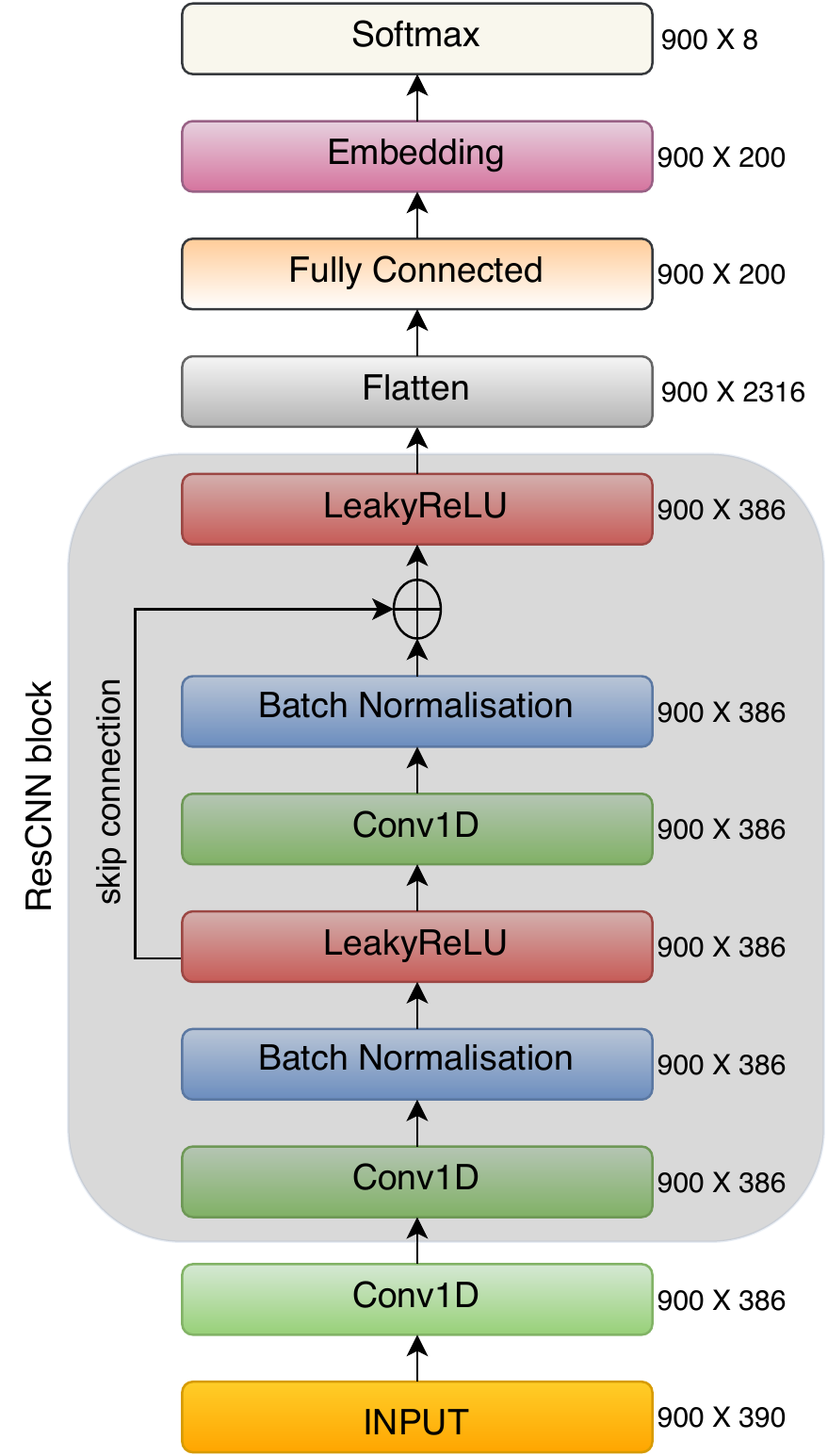}	
			\caption{Schematic description of proposed architecture.\vspace{-.2in}}
			\label{fig:archi}
		\end{center}
	\end{figure}   

	ResNet with 11 layers (and two residual blocks) performed best among these networks. It has been chosen as the suitable architecture for the proposed implementation. ResNet performs relation extraction using deep residual learning. A residual network contains skip connections among convolutional layers. It is known to perform better for large networks~\cite{li2017deep}. Fig.~\ref{fig:archi} depicts a representative diagram of the implemented ResNet architecture. The network contains the following layers: \textit{Input Layer} - to feed the input to the network in vector form; \textit{Conv1D Layers} - to build ResNet block; \textit{BatchNorm Layer} - to normalize the input and generate triplet loss; \textit{LeakyReLU Layer} - consists of the activation function to define the output of a particular layer; \textit{Flatten Layer} - to convert pooled feature map to a single column; \textit{Fully Connected Layer} - the last layer of the network that takes input from the flatten layer and \textit{Softmax Layer} - to generate cross-entropy loss. Dimensions of the layers have been shown along with them. The shape of the input is 900X390, where the first dimension 900 corresponds to the batch size, and the second dimension 390 corresponds to the feature values of an individual input speech sample.
	
	\subsection{Results and Discussion}\label{sec:results}
	The proposed system learns emotional embeddings from the model weights of the network trained on emotional speech datasets. The learned embeddings are used to recognize the speech samples of various emotions. The cosine similarity values are calculated by applying the Euclidean dot product function onto the embeddings represented in the hyperplane. Triplet loss function is used for model training. It aims to minimize the angles among the representations of the speech of similar emotions and maximize the angles among the representations of the speech of different emotions.  The computed angles are utilized to classify a new speech sample into its appropriate emotion class. The following sections discuss the learned embeddings, the angles among their projections, and the emotion classification performance.\vspace{.02in}
	
	\subsubsection{Emotion Embeddings Visualization}
\begin{figure}[!t]
		\centering
		\captionsetup{justification=centering}
		\subfloat[RAVDESS dataset]
		{\includegraphics[width=0.46\textwidth]{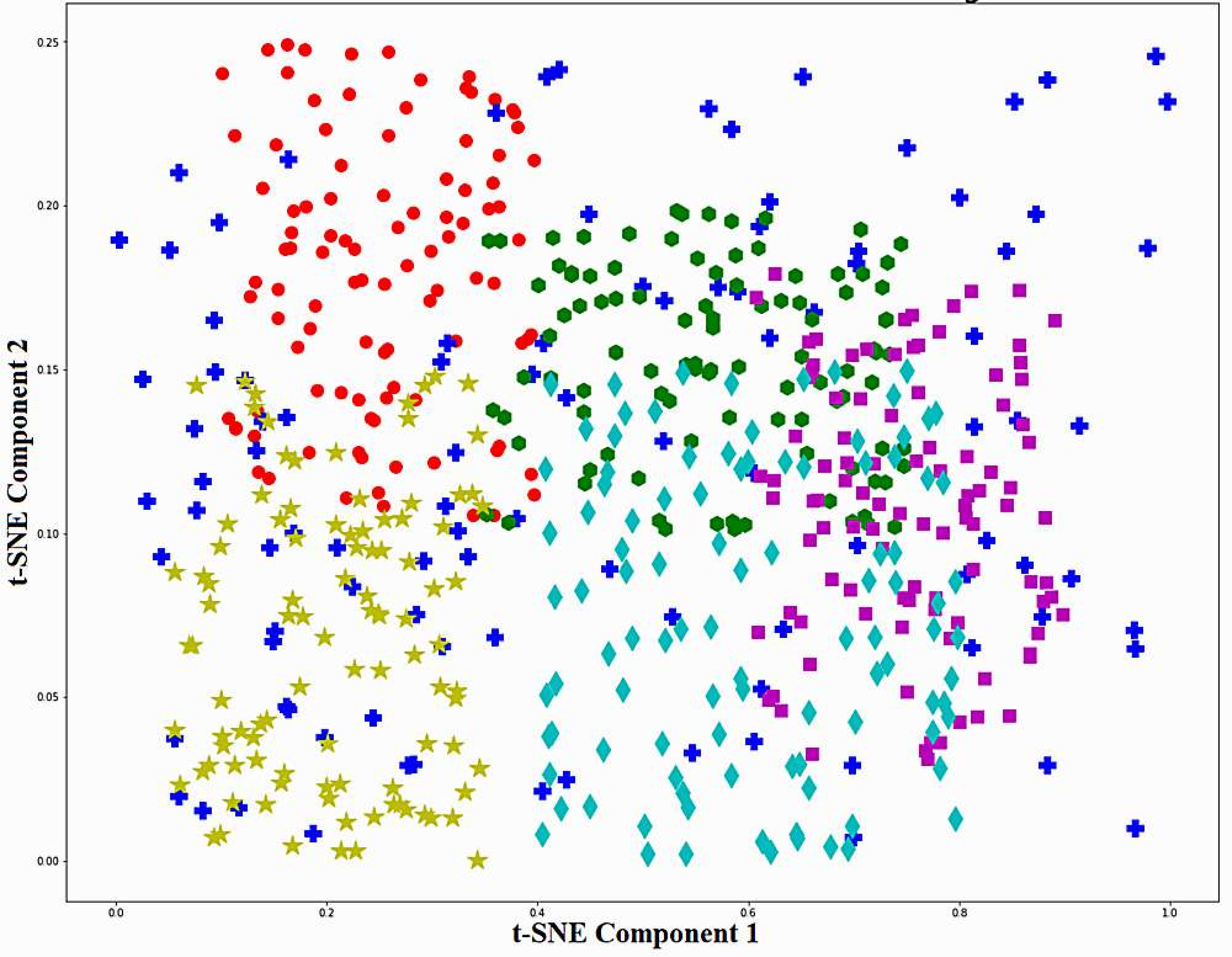}\label{fig:f1}}
		\hspace{1cm}
		\subfloat[IEMOCAP dataset]
		{\includegraphics[width=0.46\textwidth]{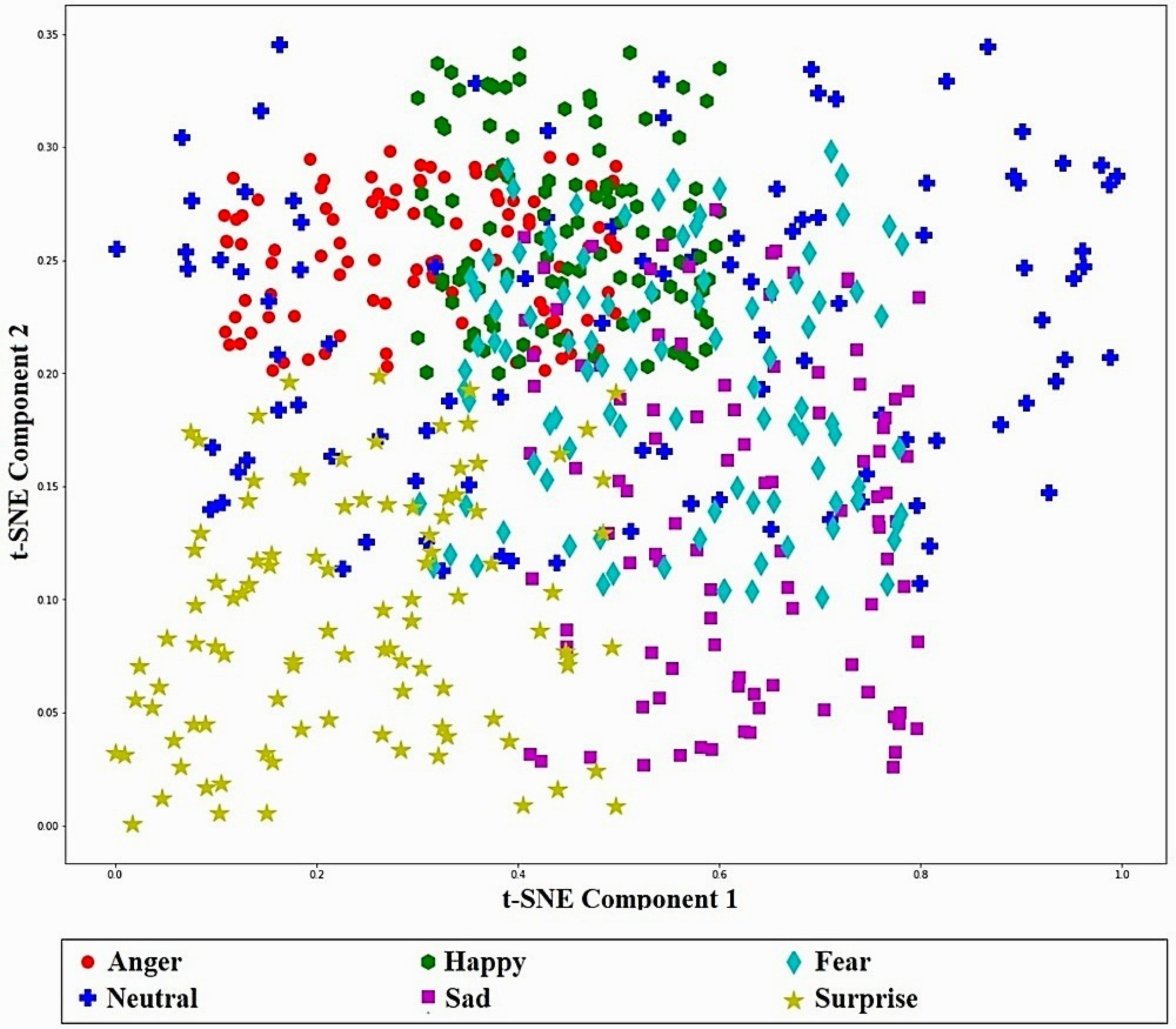}\label{fig:f2}}
		\caption{t-SNE Visualization of Emotion Embeddings.}
		\label{Emb}
	\end{figure}

	The embedding plots for six important emotions, i.e., anger, neutral, happy, sad, fear, and surprise, have been drawn in Fig.~\ref{Emb}. These are the common emotion labels for RAVDESS and IEMOCAP dataset. Hence, they are selected for the embedding plots for comparative understanding. The experimentally obtained embeddings are visualized using t-distributed stochastic neighbor embedding (t-SNE) visualization method.\vspace{.02in}
	
	\subsubsection{Emotion Classification}\label{sec:classification}
	The proposed approach has been validated for RAVDESS and IEMOCAP emotional speech dataset. The angles among the embeddings for RAVDESS and IEMOCAP are described in Table~\ref{tab:angleRAVDESS} and Table~\ref{tab:angleIEMOCAP}. The angles for a given speech utterance is checked with all the emotional classes. It is classified into the class with which it makes the least angle.\vspace{.02in}
	
	\begin{table*}[]
	\centering
	{\fontsize{8}{10}\selectfont
	\caption{Angles among various emotional embedding vectors for RAVDESS dataset}\vspace{-.15cm}
	\label{tab:angleRAVDESS}
	\resizebox{.65\textwidth}{!}
		{%
		\begin{tabular}{@{}lcccccccc@{}} \toprule
			\textbf{}         & \textbf{neutral} & \textbf{calm}   & \textbf{happy}  & \textbf{sad}    & \textbf{angry}  & \textbf{fearful} & \textbf{disgust} & \textbf{surprise} \\ \midrule
			\textbf{neutral}  & 0.37$^{\circ}$   & 30.75$^{\circ}$ & 60.34$^{\circ}$ & 89.78$^{\circ}$ & 69.73$^{\circ}$ & 62.63$^{\circ}$  & 80.28$^{\circ}$  & 77.27$^{\circ}$   \\
			\textbf{calm}     &                  & 0.51$^{\circ}$  & 78.91$^{\circ}$ & 65.01$^{\circ}$ & 64.61$^{\circ}$ & 66.37$^{\circ}$  & 89.93$^{\circ}$  & 71.71$^{\circ}$   \\
			\textbf{happy}    &                  &                 & 0.03$^{\circ}$  & 25.75$^{\circ}$ & 83.22$^{\circ}$ & 66.30$^{\circ}$  & 62.76$^{\circ}$  & 64.39$^{\circ}$   \\
			\textbf{sad}      &                  &                 &                 & 0.35$^{\circ}$  & 60.95$^{\circ}$ & 61.52$^{\circ}$  & 63.69$^{\circ}$  & 84.88$^{\circ}$   \\
			\textbf{angry}    &                  &                 &                 &                 & 0.63$^{\circ}$  & 84.97$^{\circ}$  & 61.23$^{\circ}$  & 83.40$^{\circ}$   \\
			\textbf{fearful}  &                  &                 &                 &                 &                 & 0.03$^{\circ}$   & 67.89$^{\circ}$  & 67.06$^{\circ}$   \\
			\textbf{disgust}  &                  &                 &                 &                 &                 &                  & 1.64$^{\circ}$   & 83.63$^{\circ}$ \\
			\textbf{surprise} &                  &                 &                 &                 &                 &                  &                  & 0.50$^{\circ}$  \\ \bottomrule  
		\end{tabular}%
		}
	} 
	\end{table*}
	
	\begin{table*}[]
	\centering
	{\fontsize{8}{10}\selectfont
	\caption{Angles among various emotional embedding vectors for IEMOCAP dataset}\vspace{-.15cm}
	\label{tab:angleIEMOCAP}
	\resizebox{.87\textwidth}{!}
		{
		\begin{tabular}{@{}lccccccccc@{}}
			\toprule
			\textbf{}            & \multicolumn{1}{l}{\textbf{anger}} & \multicolumn{1}{l}{\textbf{sadness}} & \multicolumn{1}{l}{\textbf{happiness}} & \multicolumn{1}{l}{\textbf{neutral}} & \multicolumn{1}{l}{\textbf{excitement}} & \multicolumn{1}{l}{\textbf{surprise}} & \multicolumn{1}{l}{\textbf{fear}} & \multicolumn{1}{l}{\textbf{disgust}} & \multicolumn{1}{l}{\textbf{frustration}} \\ \midrule
			\textbf{anger}       & 4.51$^{\circ}$                     & 51.97$^{\circ}$                      & 78.08$^{\circ}$                        & 53.39$^{\circ}$                      & 56.82$^{\circ}$                         & 79.20$^{\circ}$                       & 52.68$^{\circ}$                   & 62.32$^{\circ}$                      & 84.85$^{\circ}$                          \\
			\textbf{sadness}     &                                    & 0.96$^{\circ}$                       & 87.23$^{\circ}$                        & 70.54$^{\circ}$                      & 29.96$^{\circ}$                         & 52.28$^{\circ}$                       & 67.51$^{\circ}$                   & 85.14$^{\circ}$                      & 73.09$^{\circ}$                          \\
			\textbf{happiness}   &                                    &                                      & 0.99$^{\circ}$                         & 67.23$^{\circ}$                      & 53.74$^{\circ}$                         & 60.81$^{\circ}$                       & 77.82$^{\circ}$                   & 50.09$^{\circ}$                      & 77.06$^{\circ}$                          \\
			\textbf{neutral}     &                                    &                                      &                                        & 1.12$^{\circ}$                       & 56.43$^{\circ}$                         & 72.48$^{\circ}$                       & 61.06$^{\circ}$                   & 45.56$^{\circ}$                      & 82.68$^{\circ}$                          \\
			\textbf{excitement}  &                                    &                                      &                                        &                                      & 0.87$^{\circ}$                          & 50.74$^{\circ}$                       & 68.97$^{\circ}$                   & 88.14$^{\circ}$                      & 39.29$^{\circ}$                          \\
			\textbf{surprise}    &                                    &                                      &                                        &                                      &                                         & 2.66$^{\circ}$                        & 83.84$^{\circ}$                   & 78.18$^{\circ}$                      & 77.65$^{\circ}$                          \\
			\textbf{fear}        &                                    &                                      &                                        &                                      &                                         &                                       & 0.77$^{\circ}$                    & 32.51$^{\circ}$                      & 83.56$^{\circ}$                          \\
			\textbf{disgust}     &                                    &                                      &                                        &                                      &                                         &                                       &                                   & 0.73$^{\circ}$                       & 73.84$^{\circ}$                          \\
			\textbf{frustration} &                                    &                                      &                                        &                                      &                                         &                                       &                                   &                                      & 2.44$^{\circ}$                           \\ \bottomrule
		\end{tabular}%
		}\vspace{-.07in}
	} 
	\end{table*}
	
	The proposed approach showed an emotion recognition accuracy of \textbf{91.67}\% for RAVDESS dataset and \textbf{64.44}\% for IEMOCAP dataset. Here, `Un-weighted Accuracy' has been considered, which is defined as total correct predictions over total instances. As discussed in Tables~\ref{tab:sota_ravdess} and~\ref{tab:sota_iemocap}, the proposed approach has demonstrated comparable performance to the benchmark results. It is to be noted that the performance for RAVDESS and IEMOCAP datasets has been compared for different methods because not all of the available state-of-the-art methods included both datasets in their experiments.\vspace{.02in}

	\begin{table}[]
	\centering
	{\fontsize{8}{11}\selectfont
		\caption{Result comparison for RAVDESS dataset}\vspace{-.15cm}
		\label{tab:sota_ravdess}
		\resizebox{.49\textwidth}{!}
		{
			\begin{tabular}{@{}llc@{}}
				\toprule
				\textbf{Method}              & \textbf{Author}   & \textbf{Accuracy} \\ \midrule
				\multicolumn{2}{c}{\textit{Proposed Method}}     & \textbf{91.67\%}  \\
				Convolutional Neural Network & M. G. Pinto~\cite{marcogdepinto}             & 91.53\%           \\
				Artificial Neural Network    & K. Tomba et al.~\cite{tomba2018stress}      & 89.16\%           \\
				Multi Task Hierarichel SVM   & B. Zhang et al.~\cite{zhang2015recognizing} & 83.15\%           \\
				Bagged Ensemble of SVMs      & A. Bhavan et al.~\cite{bhavan2019bagged}    & 75.69\%           \\
				Convolutional Neural Network & D. Issa et al.~\cite{issa2020speech}        & 71.61\%           \\ \bottomrule
			\end{tabular}%
		}
	} \vspace{-.1in}
	\end{table}
	
	\begin{table}[]
	\centering
	{\fontsize{8}{11.2}\selectfont
		\caption{Result Comparison for IEMOCAP dataset}\vspace{-.15cm}
		\label{tab:sota_iemocap}
		\resizebox{.49\textwidth}{!}
		{%
			\begin{tabular}{@{}llc@{}}
				\toprule
				\textbf{Method}       & \textbf{Author}                                   & \textbf{Accuracy}                 \\ \midrule
				RNN + Attention       & N. Majunder~\cite{majumder2019dialoguernn}        & 64.50\%                           \\
				\multicolumn{2}{c}{\textit{Proposed Method}}  							   & \textbf{64.44\%} 			 	   \\
				Memory Network        & D. Hazarika et al.~\cite{hazarika2018icon}         & 63.50\%                           \\
				CNN + Mel Filterbanks & Z. Aldeneh and E. Provost~\cite{aldeneh2017using} & 61.80\%                           \\
				Memory Network        & S. Poria et al.~\cite{hazarika2018conversational}  & 56.13\%                           \\
				CNN + LSTM            & J. Zhao~\cite{zhao2019speech}                     & 52.14\%                           \\ \bottomrule
			\end{tabular}%
		}
	}\vspace{-.1in}
	\end{table}
	
	\subsubsection{Inter and Intra Class Standard Deviation}
\begin{figure}[]
		\begin{center}
			\captionsetup{justification=centering}
			\includegraphics[width=0.47\textwidth]{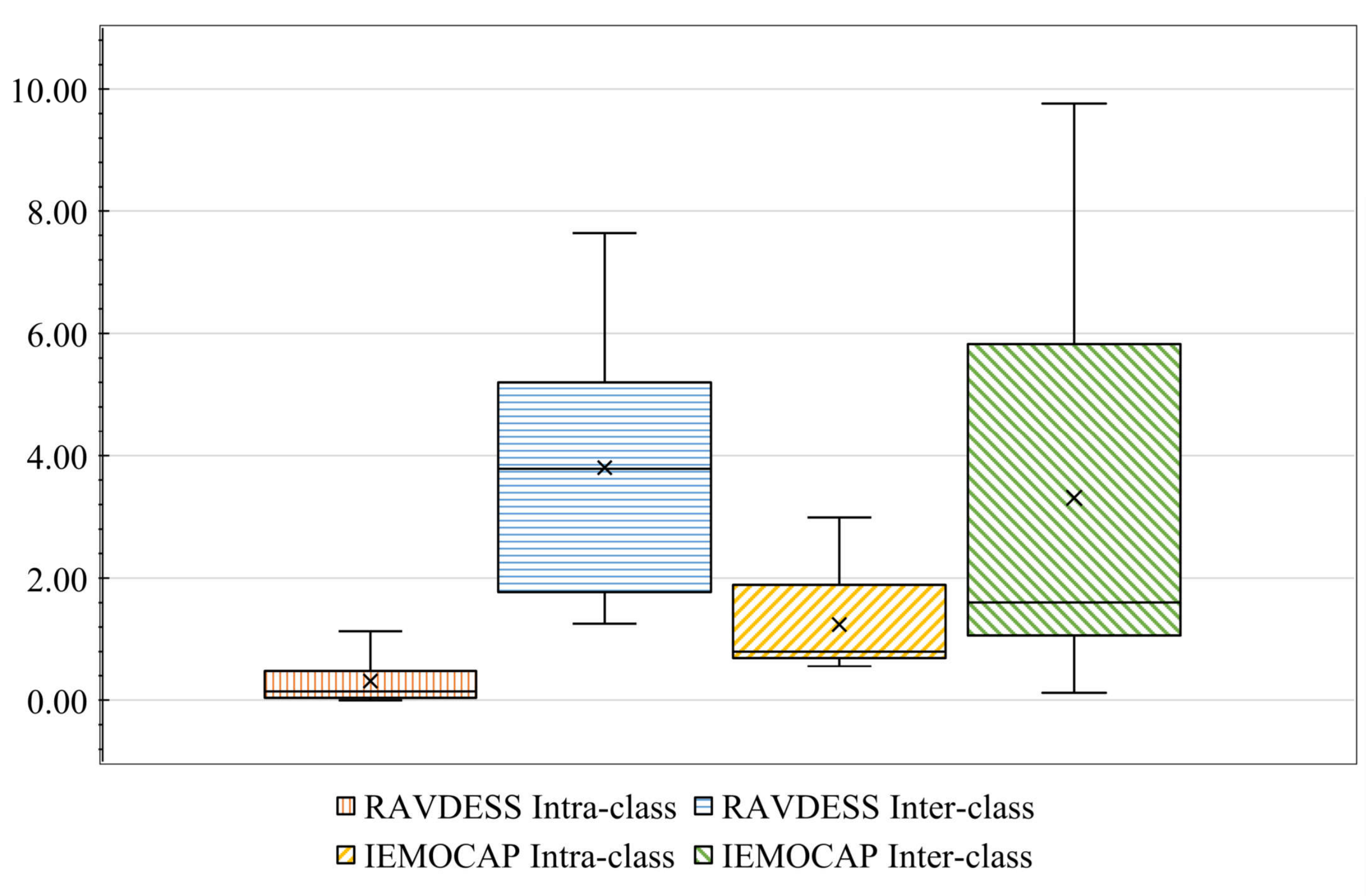}	
			\caption{Boxplots of deviation in predicted angles.\vspace{-.2in}}
			\label{bplot}
		\end{center}
	\end{figure}

	A given speech utterance is classified into the emotion class with which it makes the least angle. As detailed in Tables~\ref{tab:var1} and~\ref{tab:var2}, deviation in the predicted angles among emotion embedding vectors is observed among similar (intra-class) and different (inter-class) emotion classes. The average intra-class angles are 0.51 and 1.68 for RAVDESS and IEMOCAP. While average inter-class angles are 68.41 and 65.88, respectively. Fig.~\ref{bplot} presents the boxplots for the deviation values. Deviation in intra-class angles affects SER accuracy significantly as it denotes the projections of the embeddings of the speech utterances in the hyperplane. Lesser deviation in the intra-class angles for RAVDESS corresponds to the closeness in their projections and more accurate SER predictions as compared to IEMOCAP.
	
	\begin{table}[]
	\centering
	{
		\fontsize{8}{10.2}\selectfont
		\caption{Std. Dev. analysis for RAVDESS dataset}\vspace{-.15cm}
		\label{tab:var1}
		\resizebox{.45\textwidth}{!}
		{
			\begin{tabular}{@{}lcc@{}}
				\toprule
				\textbf{} & \textbf{Intra-class deviation} & \textbf{Inter-class deviation} \\ \midrule
				neutral   & 0.14                           & 2.11                           \\
				calm      & 0.00                           & 1.66                           \\
				happy     & 0.48                           & 5.32                           \\
				sad       & 0.15                           & 4.87                           \\
				angry     & 0.12                           & 3.21                           \\
				fearful   & 0.47                           & 1.26                           \\
				disgust   & 1.13                           & 4.36                           \\
				surprise  & 0.01                           & 7.64                           \\ \bottomrule
			\end{tabular}%
		}
	} \vspace{-.1in}
	\end{table}

\begin{table}[]
	\centering
	{\fontsize{8}{10.2}\selectfont
	\caption{Std. Dev. analysis for IEMOCAP dataset}\vspace{-.15cm}
	\label{tab:var2}
	\resizebox{.45\textwidth}{!}
	{
		\begin{tabular}{@{}lcc@{}}
			\toprule
			\textbf{}   & \textbf{Intra-class deviation} & \textbf{Inter-class deviation} \\ \midrule	
			anger       & 2.84                           & 0.96                           \\
			sadness     & 0.71                           & 1.16                           \\
			happiness   & 0.68                           & 3.13                           \\
			neutral     & 0.56                           & 1.60                           \\
			excitement  & 0.80                           & 9.76                           \\
			surprise    & 0.99                           & 3.52                           \\
			fear        & 0.91                           & 0.12                           \\
			disgust     & 0.94                           & 1.41                           \\
			frustration & 0.77                           & 8.13                           \\ \bottomrule
		\end{tabular}%
	}} 
\end{table}
	
	As observed in Fig.~\ref{Emb}, the emotion embeddings for RAVDESS dataset are more clearly defined than IEMOCAP dataset. It is also noted that some overlap has been observed between the embeddings of `sad' and `fear', both of which are negative-valence emotions. On the other hand, both `anger' and `surprise' are emotions with high intensity. They also showed minor overlap between their embeddings. Another observation is that the network converged faster during the training when softmax pre-training was used along with triplet loss training.
	
	\section{Conclusions and future work}\label{sec:conclusion}	
	In this paper, an end-to-end emotion embedding system has been proposed to learn the emotional patterns from speech samples in the form of an embedding matrix. It projects the mappings of speech information onto a hyperplane where triplet loss is used as a loss function to learn the similarities among various emotions based on cosine similarity. The emotion embedding matrix thus prepared has been used for speech emotion recognition and it demonstrated comparable recognition results to the benchmarks for two different datasets. The proposed system has classified the emotions for RAVDESS dataset with an accuracy of 91.67\%, while 64.44\% accuracy has been observed for IEMOCAP dataset.\vspace{.02in}
	
	Experiments for various neural architectures to automatically learn the acoustic features were performed, and finally, ResNet was implemented, which demonstrated better performance in finding relational emotion patterns for speech samples. The current implementation requires checking of angles for each speech utterance with each emotion class. This process can be optimized with the aim of reducing training time, computational requirements, and model size In the future, it is also aimed to use the learned embeddings for other speech processing tasks such as emotional speech synthesis.
	
	\section*{Acknowledgements}
	This research was supported by the Ministry of Human Resource Development (MHRD) INDIA with reference grant number: 1-3146198040.\\
	
	{\small
		\bibliographystyle{IEEEtran}
		\bibliography{root}
	}
	
\end{document}